\documentclass[twocolumn,apj,appendixfloats,numberedappendix]{emulateapj}
\usepackage{hyperref}
\usepackage{amsfonts,amsmath,amssymb,amsthm}
\usepackage{longtable}
\usepackage{graphicx}
\usepackage{tabularx}
\usepackage{color}
\usepackage[T1]{fontenc}
\usepackage{lmodern}
\usepackage[right]{rotating}
\usepackage{fancyref}
\usepackage{tabularx}
\usepackage{textcomp}
\usepackage{enumerate}

\newcolumntype{L}[1]{>{\raggedright\arraybackslash}p{#1}}
\newcolumntype{C}[1]{>{\centering\arraybackslash}p{#1}}
\newcolumntype{R}[1]{>{\raggedleft\arraybackslash}p{#1}}
\newcolumntype{Y}{>{\centering\arraybackslash}X}

\newcommand{\vect}[1]{{\boldsymbol{#1}}}

\shorttitle{ION-SCALE SPECTRAL BREAK: 2D HYBRID SIMULATIONS}
\shortauthors{FRANCI ET AL.}

\begin{document}

\title{Plasma beta dependence of the ion-scale spectral break of solar wind turbulence: high-resolution 2D hybrid simulations}

\author{Luca Franci\altaffilmark{1,2}, Simone Landi\altaffilmark{1,3}, Lorenzo Matteini\altaffilmark{4}, Andrea Verdini\altaffilmark{5}, Petr Hellinger\altaffilmark{6}}
\altaffiltext{1}{Dipartimento di Fisica e Astronomia, Universit\`a degli Studi di Firenze, Largo E. Fermi 2, I-50125 Firenze, Italy}
\altaffiltext{2}{INFN - Sezione di Firenze, Via G. Sansone 1, I-50019 Sesto F.no (Firenze), Italy}
\altaffiltext{3}{INAF - Osservatorio Astrofisico di Arcetri, Largo E. Fermi 5, I-50125 Firenze, Italy}
\altaffiltext{4}{Space and Atmospheric Physics Group, Imperial College London, London SW7 2AZ, UK}
\altaffiltext{5}{LESIA - Observatoire de Paris-Meudon, 5, place J. Janssen, F-92195 Meudon Cedex, France}
\altaffiltext{6}{Astronomical Institute, CAS, Bocni II/1401, CZ-14100 Prague, Czech Republic}
\date{\today}

\begin{abstract}
We investigate properties of the ion-scale spectral break of solar
wind turbulence by means of two-dimensional high-resolution hybrid
particle-in-cell simulations.  We impose an initial ambient magnetic
field perpendicular to the simulation box and add a spectrum of
in-plane, large-scale, magnetic and kinetic fluctuations.  We perform
a set of simulations with different values of the plasma $\beta$,
distributed over three orders of magnitude, from $0.01$ to $10$.  In
all the cases, once turbulence is fully developed, we observe a
power-law spectrum of the fluctuating magnetic field on large scales
(in the inertial range) with a spectral index close to $-5/3$, while
in the sub-ion range we observe another power-law spectrum with a
spectral index systematically varying with $\beta$ (from around $-3.6$
for small values to around $-2.9$ for large ones).  The two ranges are
separated by a spectral break around ion scales. The length scale at
which this transition occurs is found to be proportional to the ion
inertial length, $d_i$, for $\beta \ll 1$ and to the ion gyroradius,
$\rho_i = d_i\sqrt{\beta}$, for $\beta \gg 1$, i.e., to the larger
between the two scales in both the extreme regimes. For intermediate
cases, i.e., $\beta \sim 1$, a combination of the two scales is
involved.  We infer an empiric relation for the dependency of the
spectral break on $\beta$ that provides a good fit over the whole
range of values.  We compare our results with in situ observations in
the solar wind and suggest possible explanations for such a behavior.
\end{abstract}
\LTcapwidth=\columnwidth 


\maketitle

\section{Introduction}
\label{sec:introduction}
The solar wind is an exceptional laboratory for plasma astrophysics
thanks to spacecraft in situ observations. One of the best established
observational results is a ubiquitous presence of a broadband range of
electromagnetic fluctuations interpreted as a turbulent cascade
connecting the fluid motion on large scales to small-scale kinetic
fluctuations at particle scales \citep{Bruno_Carbone_2013}.  At large
scales, turbulent fluctuations exhibit properties consistent with
magnetohydrodynamics (MHD) turbulence
\citep[e.g.,][]{Bavassano_al_1982, Marsch_Tu_1990,
  Grappin_al_1990}. Approaching particle characteristic scales, a
transition to a different, kinetic, regime of the turbulence is
observed. This regime is characterized by a steepening of the magnetic
field spectrum, followed by a further steepening at electron scales
\citep{Alexandrova_al_2009, Sahraoui_al_2013}. A clear change in the
magnetic field spectral slope is observed between the MHD and the
sub-ion range \citep[e.g.,][]{Leamon_al_1998, Bruno_al_2014,
  Lion_al_2016}, going from a Kolmogorov-like scaling with a spectral
index of $-5/3$ to a steeper power law, phenomenologically consistent
with a spread of the spectral index around a typical value of
$-2.8$. 

In the solar wind, the transition between MHD and kinetic turbulence
occurs close to the convected characteristic ion scales, namely the
ion inertial length, $d_i$, and the ion Larmor radius,
$\rho_i$. However, it is not straightforward to conclude from
observations which of the two scales (or what kind of their
combination) is associated to the spectral change and, consequently,
which are the physical processes governing the transition and the
cascade at sub-ion scales. The main reason is that the two scales are
very close to each other under typical solar wind conditions, since
$\rho_i=\sqrt{\beta_i}\,d_i$ and the ion plasma beta, $\beta_i$, is of
the order of 1 in the vicinity of 1 astronomical unit (au) (see
Sec. \ref{sec:setup} for the definitions of $d_i$, $\rho_i$, and
$\beta_i$). Moreover, the radial evolution of the spectral break does
not suggest any firm evidence in favor of any of the two scales
\citep{Perri_al_2010, Bourouaine_al_2012, Bruno_Trenchi_2014}.

A recent study by \cite{Chen_al_2014} investigated extreme regimes of
$\beta_i$ measured by the WIND spacecraft at 1 au, and provided a
clear evidence of a beta dependence of the ion-scale break in solar
wind turbulence. The main result of this study is that there is not a
single scale associated to the spectral break for all values of
$\beta_i$.  Indeed, the spatial ion scale associated to the spectral
break is observed to be always the largest of the two, namely $d_i$
for $\beta_i\ll1$, and $\rho_i$ for $\beta_i\gg1$. This suggests that
the first relevant scale encountered by the turbulent fluctuations is
the one that determines the transition and the properties observed in
the sub-ion regime.

\begin{table*}[!t]
\begin{tabularx}{\textwidth}{c|YYY|YYY|YYY} 
\hline \hline
RUN & $B^{\textrm{rms}} (B_0)$ & $\beta$ & $k^{\textrm{inj}} d_i$ & $\Delta x \; (d_i)$ & $\eta \; (4\pi/\omega_p)$ & ppc & $\alpha_1$ & $\alpha_2$ & $k^{b}_\perp d_i$ \\
\hline \hline 
1  & $0.06$ & $1/100$ & 0.2  & 0.125 & $5 \times 10^{-4}$ & $1000$ & -1.71 & -3.52 & 2.89 \\
2  & $0.12$ & $1/32$ & 0.2  & 0.125 & $5 \times 10^{-4}$ & $1000$  & -1.64 & -3.53 & 3.47 \\
3  & $0.24$ & $1/16$ & 0.2  & 0.125 & $5 \times 10^{-4}$ & $1000$  & -1.68 & -3.22 & 3.46 \\
4  & $0.24$ & $1/8$  & 0.2  & 0.125 & $5 \times 10^{-4}$ & $2000$  & -1.71 & -3.22 & 3.41 \\
5  & $0.24$ & $1/4$  & 0.2  & 0.125 & $5 \times 10^{-4}$ & $4000$  & -1.71 & -3.06 & 3.01 \\
6  & $0.24$ & $1/2$  & 0.2  & 0.125 & $5 \times 10^{-4}$ & $8000$  & -1.65 & -3.00 & 2.55 \\
7  & $0.24$ & $1$    & 0.2  & 0.125 & $5 \times 10^{-4}$ & $12000$ & -1.55 & -2.87 & 2.06 \\
8  & $0.24$ & $2$    & 0.2  & 0.125 & $5 \times 10^{-4}$ & $16000$ & -1.54 & -2.87 & 1.90 \\
9  & $0.24$ & $4$    & 0.2  & 0.125 & $5 \times 10^{-4}$ & $16000$ & -1.65 & -2.91 & 1.59 \\
10 & $0.48$ & $6$    & 0.05 & 0.25  & $1 \times 10^{-3}$ & $8000$  & -1.75 & -2.91 & 1.09 \\
11 & $0.48$ & $8$    & 0.05 & 0.25  & $1 \times 10^{-3}$ & $8000$  & -1.71 & -3.01 & 1.10 \\
12 & $0.48$ & $10$   & 0.05 & 0.25  & $1 \times 10^{-3}$ & $8000$  & -1.70 & -2.99 & 1.01 \\
\hline \hline
\end{tabularx}
\caption{List of Simulations and Their Relevant Parameters}
\label{tab:simulations_beta}
\end{table*}

Numerical simulations retaining ion kinetic effects
\citep[e.g.][]{Howes_al_2011, Parashar_al_2010, Passot_al_2014,
  Servidio_al_2012, Valentini_al_2014, Cerri_al_2016} are able to
capture some of the phenomenology of the ion scale transition, leading
to magnetic spectra with a steeper slope at sub-ion scales.  In
particular, high-resolution two-dimensional (2D) hybrid
particle-in-cell (PIC) simulations by \cite{Franci_al_2015a,
  Franci_al_2015b} successfully reproduce many of the observational
characteristics of the transition of the turbulent cascade from MHD to
kinetic scales, including a quantitative agreement of spectral slopes
and compressibility and energy ratios. These works considered only one
intermediate beta regime, $\beta_i=0.5$. Different values of $\beta_i$
were already investigated by means of a hybrid Vlasov-Maxwell model
\citep{Servidio_al_2014}, mainly focusing on the particle anisotropy
associated to different plasma conditions. More recently, employing
the same approach, \cite{Cerri_al_2016} studied the dependence of the
physics of subproton-scale kinetic turbulence on $\beta_i$ by
exploring three particular cases, i.e., $\beta_i=0.2$, $1$, and $5$.
They observe a dominance of magnetosonic/whistler fluctuations in the
low-beta case and of kinetic Alfv\'en waves (KAWs) in the high-beta
case. The numerical study of \cite{Cerri_al_2016} is not, however,
directed to the ion spectral break as such.

In this paper we investigate the properties of turbulence and its
transition from the MHD to the sub-ion regime over a very wide range
of plasma betas, similar to that in \cite{Chen_al_2014}.  We present
a parameter study on $\beta_i$ performed by means of twelve
high-resolution 2D hybrid particle-in-cell simulations, focussing on the
scale associated to the ion break and the steepening of the magnetic
field spectrum at sub-ion scales.  For extreme $\beta_i$ we recover
the observational results of \cite{Chen_al_2014}, whereas for
intermediate cases a combination of $d_i$ and $\rho_i$ seems to
be involved.  We infer an empiric relation of the break scale as a
function of $\beta_i$ that provides a good fit over the whole range of
values.  Finally, we offer a physical interpretation of the
observed phenomena.

\section{Numerical setup and initial conditions}
\label{sec:setup}

We use the hybrid-PIC code CAMELIA (Current Advance Method Et cycLIc
leApfrog), where the electrons are considered as a massless,
charge-neutralizing fluid with a constant temperature, whereas the
ions are described by a particle-in-cell model and are advanced by a
Boris scheme. A detail description of the model equations can be found
in \citet{Matthews_1994}.  Units of space and time are
the ion (proton) inertial length, $d_i = c/\omega_{\rm p}$
($\omega_{\rm p}$ being the proton plasma frequency), and the inverse
proton gyrofrequency, $\Omega^{-1}_{\rm p}$, respectively. The plasma
beta for a given plasma species, protons or electrons, is $\beta_{\rm
  p,e} = 8 \pi n K_B T_{\rm p,e} / B^2_0$, where $n = n_{\rm p} =
n_{\rm e}$ is the number density, assumed to be equal for protons and
electrons, $B_0$ the ambient magnetic field, $K_{\rm B}$ the
Boltzmann constant, and $T_{\rm p,e}$ the proton and electron
temperatures. For a complete definition of all quantities, please
refer to \citet{Franci_al_2015a, Franci_al_2015b}.
In this paper, we present results from twelve high-resolution
simulations with different values of the plasma beta, including the
case already presented in \citet{Franci_al_2015a,Franci_al_2015b}.
The adopted simulation box is a square grid with $2048^2$ cells in the
$(x,y)$ plane. The spatial resolution, $\Delta x= \Delta y$, and
consequently the box size, is not the same for all the simulations and
the time step for the particle advance is adjusted proportionally. All
simulations employ a few thousands of particles per cell (ppc),
corresponding to many billions of particles in the whole computational
grid.

The initial setup we employ here is the same as in
\citet{Franci_al_2015a, Franci_al_2015b}: we initialize with an
initial spectrum of magnetic and velocity fluctuations in the $(x,y)$
plane and we impose an initial ambient magnetic field, $\vect{B}_0
=B_0 \, \vect{z}$, in the perpendicular direction.  The initial
fluctuations are composed of modes having all the same amplitude and
random phases and are characterized by energy equipartition and
vanishing correlation between kinetic and magnetic fluctuations.
Their global amplitude, estimated as the root mean square value (rms)
of the total magnetic field $B$ computed over the whole simulation
domain, $B^{\textrm{rms}}$, is not the same for all simulations.  We
assume that protons are initially isotropic with a given
$\beta_i$. Electrons are also isotropic and their beta is always set
to be $\beta_{\rm e} = \beta_{\rm p}$, henceforth we will simply
denote the proton/electron plasma beta as $\beta$.  The values of
$\beta$ span three full orders of magnitude, from 0.01 to 10, so that
the ion inertial length, $d_i$, and the ion gyroradius, $\rho_i = d_i
\,\sqrt{\beta_\perp}$, are well separated at the extreme values of
$\beta$.

The main parameters of all the simulations are summarized in
Tab. \ref{tab:simulations_beta}.  In the first column we assign a
number to each run, while in the next six we report, from left to
right: the initial rms value of the perpendicular magnetic field,
$B^{\textrm{rms}}$ (in units of the ambient magnetic field, $B_0$), the
plasma beta, $\beta$, the injection scale, $k^{\textrm{inj}} d_i$,
i.e., the maximum scale of the initial fluctuations, the spatial
resolution, $\Delta x$ (in units of $d_i$), the value of the
resistivity coefficient, $\eta$ (in units of $4 \pi / \omega_{\rm
  p}$), and the number of ppc.  In the last three columns we report
the results of our analysis, which will be described in
Sec.~\ref{sec:results}.

For larger values of $\beta$, we need to employ more ppc to keep under
control the ppc-noise level at small scales, or alternatively to
increase the amplitude of the initial fluctuations,
$B^{\textrm{rms}}$.  When $\beta$ is quite low, the proton gyroradius
gets small and possibly comparable with the spatial resolution, so we
need to employ smaller grid cells and the time step must be reduced
accordingly.  A non-zero resistivity has been introduced in order to
guarantee a satisfactory conservation of the total energy, with no
claim to model any realistic physical process.  The resistivity
coefficient, $\eta$, has been fine-tuned accordingly with the
discussion presented in \citet{Franci_al_2015b}, so that the
conservation of the total energy is ensured with an accuracy of less
then $0.5$~\% for all the simulations.

\section{Results}
\label{sec:results}

\begin{figure}
\centering
\includegraphics[width=0.48\textwidth]{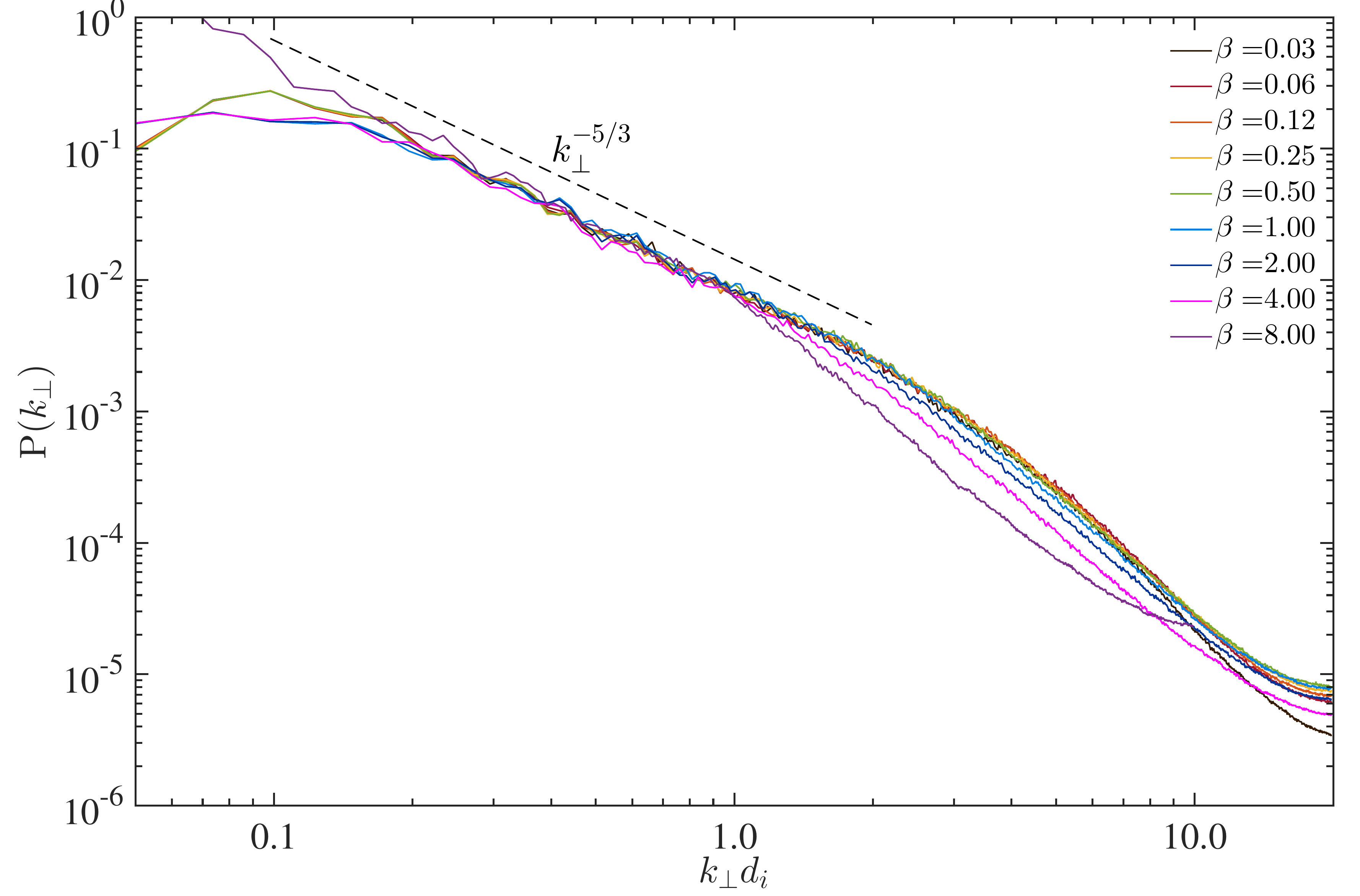}\\ 
\includegraphics[width=0.48\textwidth]{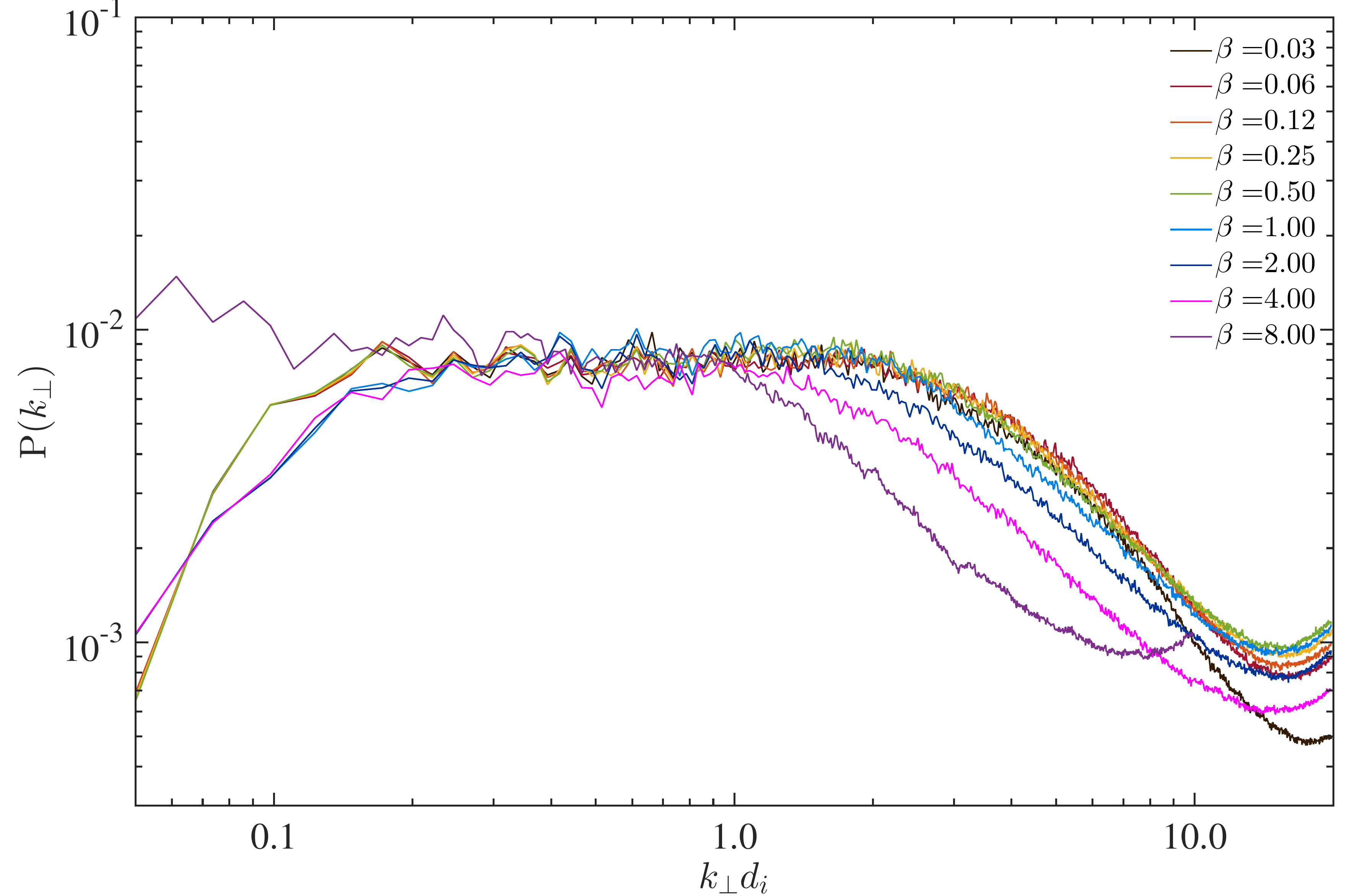}\\ 
\includegraphics[width=0.48\textwidth]{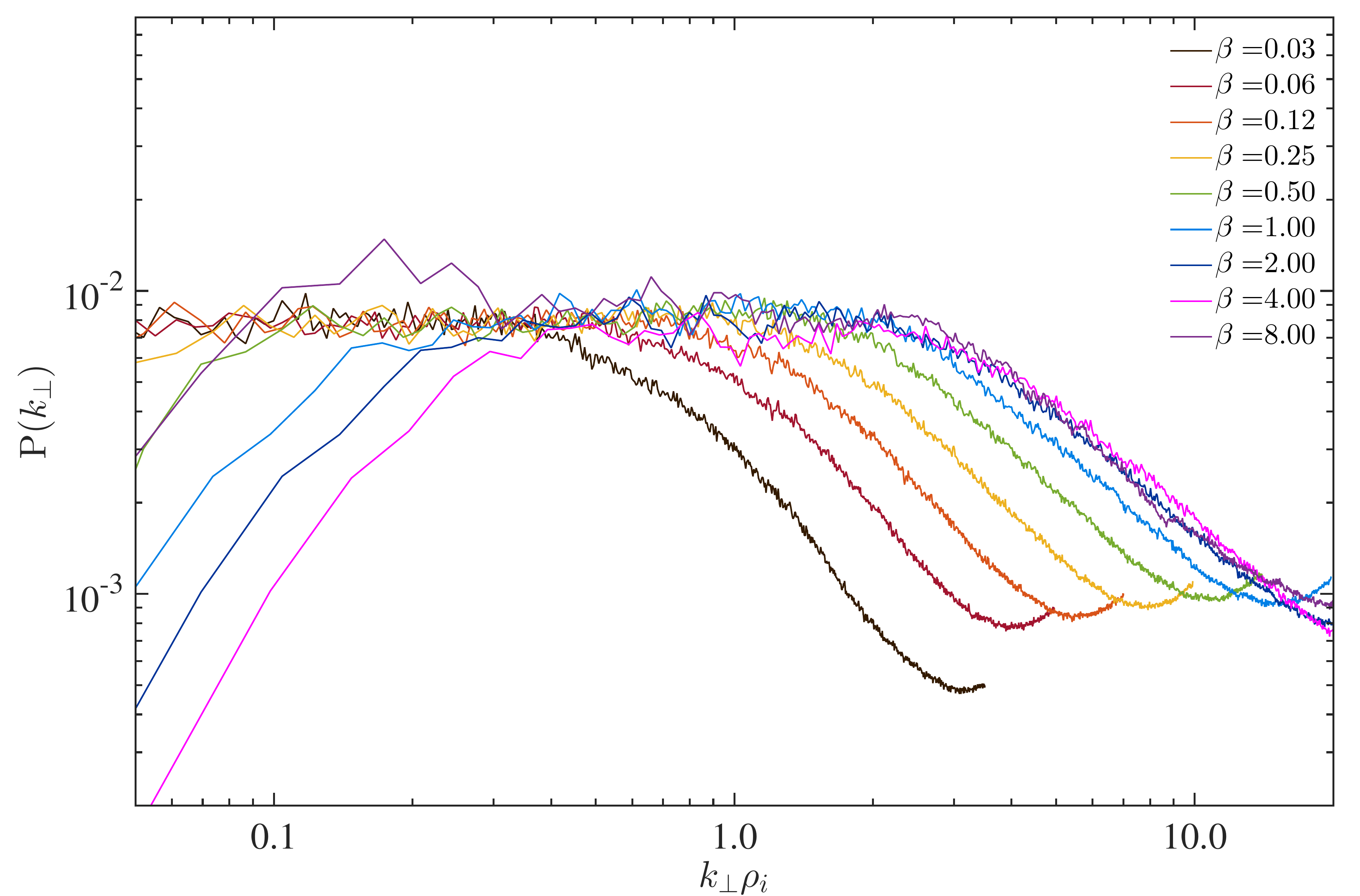}
\caption{Top panel: power spectra of magnetic fluctuations for
  different values of the plasma beta, $\beta$, versus $k_\perp
  d_i$. Middle panel: power spectra of magnetic fluctuations for
  different values of $\beta$, compensated by $k_\perp^{5/3}$, versus
  $k_\perp d_i$. Bottom panel: the same as in the middle panel, but
  versus $k_\perp \rho_i$.}
\label{fig:spectra_together}
\end{figure}

\begin{figure}
\centering 
\includegraphics[width=0.48\textwidth,trim={0 1.4cm 0 0},clip=true]{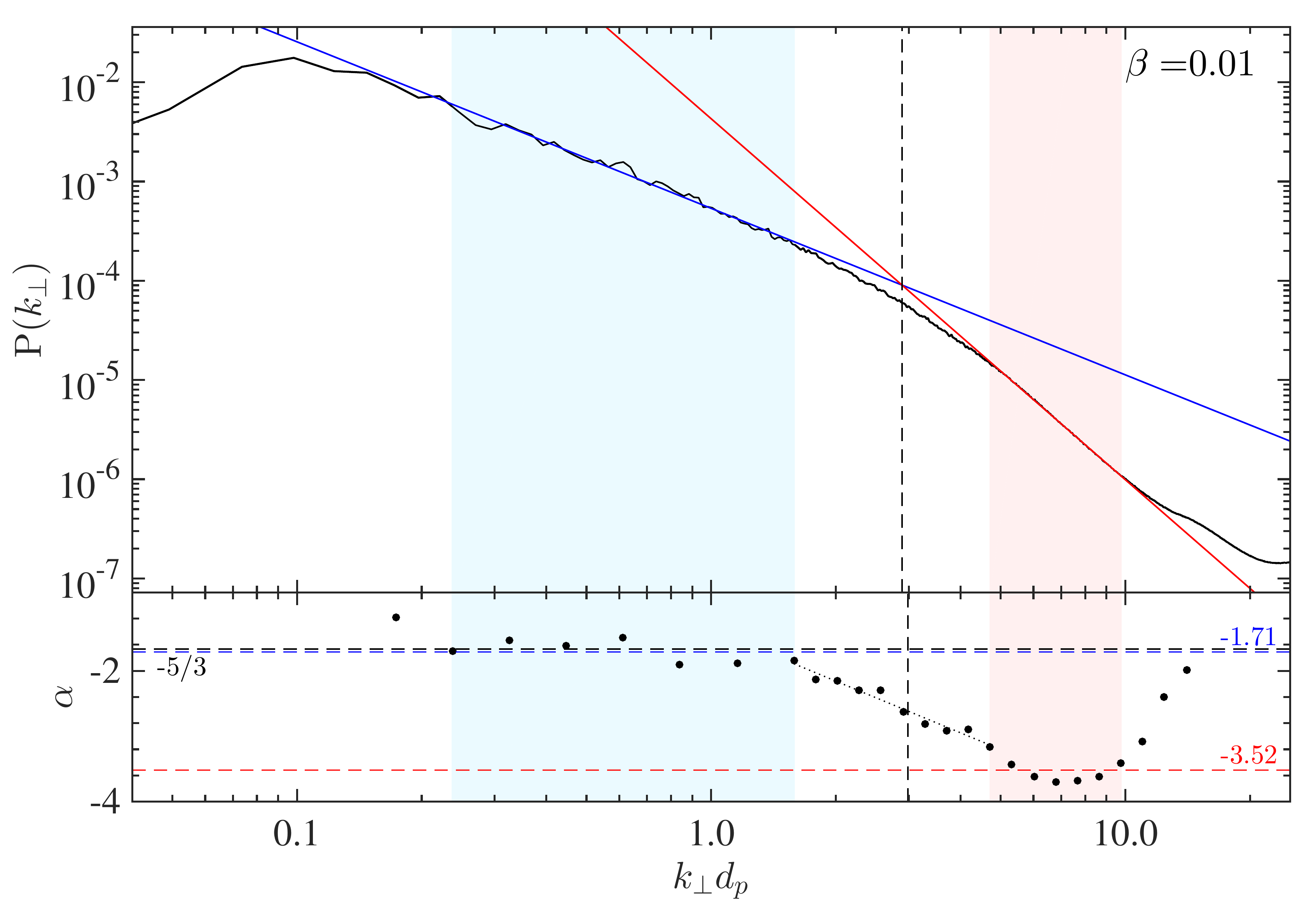}
\includegraphics[width=0.48\textwidth,trim={0 1.4cm 0 0},clip=true]{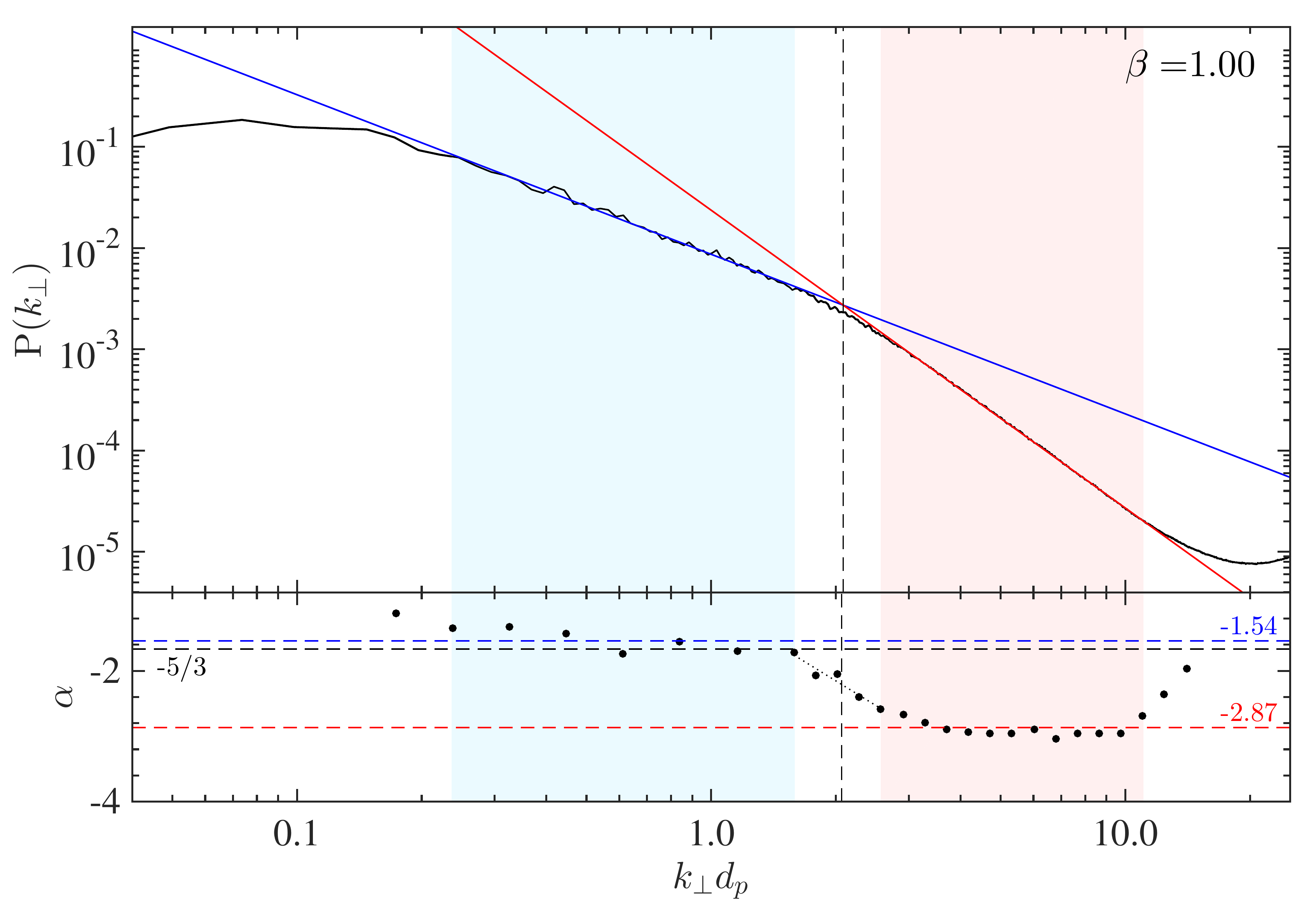}
\includegraphics[width=0.48\textwidth,trim={0 1.4cm 0 0},clip=true]{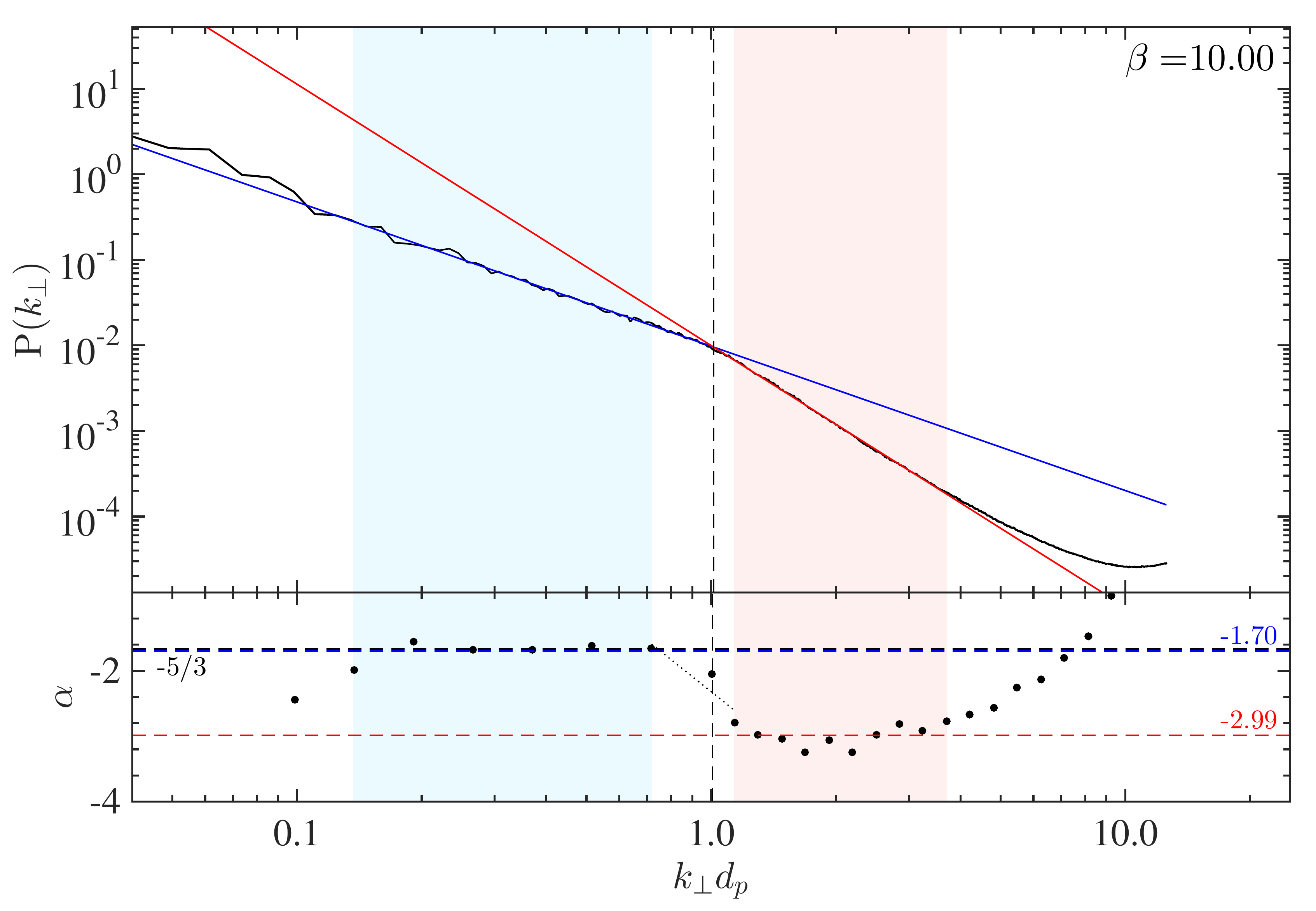}
\caption{Power spectra of magnetic fluctuations for three different
  values of the proton plasma beta representing different regimes,
  i.e., $\beta = 0.01$ (top panel), $\beta = 1$ (middle panel), and
  $\beta = 10$ (bottom panel). The light blue and light red shaded
  regions mark the intervals where the global fits of the power laws
  were performed, for the inertial and the kinetic ranges,
  respectively. In the bottom parts of each panel, the value of the
  local spectral index, $\alpha$, is also reported.}
\label{fig:spectra_singlebeta}
\end{figure}

All the quantities shown in the present paper are computed at the time
of maximum turbulent activity, i.e., at the time when the out-of-plane
component of the current density maximizes
\citep{Mininni_Pouquet_2009}. The spectral properties remain quite
stable afterwards \citep{Franci_al_2015b}. The raw data (i.e., the
magnetic field components) from which all the spectra were computed
are available online \citep{Franci_Datasets_2016}, so that all the
results presented here can be easily reproducible.

In the top panel of Fig.~\ref{fig:spectra_together} we show the power
spectrum of magnetic fluctuations for many different values of the
plasma beta, $\beta$, versus $k_\perp d_i$ (for the sake of clarity,
we decided not to include all the simulations here, but note that the
two missing extreme cases, i.e., $\beta = 0.01$ and $10$, are shown
separately in Fig.~\ref{fig:spectra_singlebeta}).
The spectra have been re-normalized to take into account the
different amplitude of the initial fluctuations, so that they have the
same power in the inertial range and can be compared more directly.
All of them exhibit a power-law behavior with a Kolmogorov-like scaling
in the inertial range (a $-5/3$ power law is drawn with a
black dashed line as a reference), a more or less smooth break at ion
scales and another power-law interval at sub-ion scales. 
The flattening of the spectra at higher wavevectors is not physical
and only due to numerical noise.  In the middle and bottom panels of
the same figure, we also report all the spectra of magnetic
fluctuations, compensated by $k_\perp^{5/3}$, as a function of
$k_\perp d_i$ and $k_\perp \rho_i$, respectively. In the middle panel,
all the spectra with low betas tend to overlap while the others don't,
meaning that the scale of the break is fixed with $d_i$ for $\beta \ll
1$.  In the bottom panel, the opposite situation holds, i.e., all the
spectra with high betas tend to overlap while the others don't,
meaning that the scale of the break is fixed with $\rho_i$ for $\beta
\gg 1$.  Therefore, Fig.~\ref{fig:spectra_together} already provides a
qualitative indication that the spectral break seems to be related to
the larger of the two scales in both regimes, i.e., $d_i$ for $\beta
\ll 1$ and $\rho_i$ for $\beta\gg 1$.  In order to quantitatively
confirm this idea, we looked at each spectrum separately and computed
the break for each of them.

In Fig.~\ref{fig:spectra_singlebeta} the spectra of the total magnetic
power is reported for three representative cases: the lowest plasma
beta, $\beta=0.01$ (top panel), the intermediate value, $\beta=1$ (middle
panel), and the highest value $\beta=10$ (bottom panel).
The shape of the ion-scale break is quite different for different
values of $\beta$: while it is quite sharp when $\beta \gg 1$ (bottom
panel of Fig.~\ref{fig:spectra_singlebeta}), it becomes smoother when
$\beta$ is low (top panel) and, in the cases with very low values
determining a length scale associated to the break is not
straightforward, since it might also depend on the criterion chosen to
define the break itself.
In order to determine such scale, we employ two different methods.
The first method is the same applied by
\citet{Chen_al_2014} and we choose it in order to directly compare
our numerical results with their observational data.  Firstly, we
compute a local power-law fit of the magnetic field spectrum over
many small intervals in the range $k_\perp d_i \in [0.15,\,15]$.
The values of the local spectral index, $\alpha$, for each simulation
are shown in the bottom part of each panel
of Fig.~\ref{fig:spectra_singlebeta}.  We consider a range of
wavevectors where $\alpha$ is close to $-5/3$ within a relative
accuracy of $\pm 20\%$ (the light blue shaded region marks its
boundaries) and we fit the values of $\alpha$ within this interval
with an horizontal line (blue dashed line in the bottom panel),
getting a value for the spectral index in the inertial range,
$\alpha_1$.
The sub-ion power-law index, $\alpha_2$, is determined in a similar way:
we select a range of $k_\perp$ where $\alpha$ is constant
within a relative accuracy of $\pm 10\%$, without assuming any
specific value a-priori, and we perform a fit 
over this interval (indicated by a light red shaded region).
Now we define the scale of the break as the wavevector at which
$\alpha$ takes a value half way between $\alpha_1$ and $\alpha_2$.  The
two spectral indices, $\alpha_1$ and $\alpha_2$, and the scale
corresponding to the spectral break, $k^{b}_\perp d_i$, are reported
in the last three columns of Tab.~\ref{tab:simulations_beta}.

In the inertial range, the power spectra of magnetic fluctuations
exhibit a Kolmogorov-like behavior for all values of $\beta$, as
already shown qualitatively in Fig.~\ref{fig:spectra_together}.  The
results of the fits for the power-law index at large scales,
$\alpha_1$, are reported in Tab.~\ref{tab:simulations_beta}. They are
all quite close to $-5/3$, which represents the mean value, and
exhibit variations of a few percent with no correlation with
$\beta$. Although departures from the Kolmogorov expectation are
actually observed in the solar wind \citep[e.g.,][]{Tessein_al_2009,
  Chen_al_2013}, in our simulations they seem to be mainly due to the
choice of the fit interval in the inertial range, which is slightly
different for each value of $\beta$ (note that the break shifts
towards larger scales for larger betas and the inertial range gets
consequently shorter).

The results of the fits for the power-law index at sub-ion scales,
$\alpha_2$, show quite larger variations, as can be seen from the
second last column of Fig.~\ref{fig:spectra_together}.  Indeed,
$\alpha_2$ is systematically less and less steep increasing the plasma
beta, ranging from around $-3.6$ for $\beta = 0.01$ until around
$-2.9$ for $\beta = 4$, although we observe a more general power-law
spectrum with a constant spectral index $\sim -2.8$ for the parallel
magnetic fluctuations instead (see Sec.~\ref{sec:conclusions}).  The
fact that the slope increases a little bit again towards $-3$ for
$\beta > 4$ is likely due to numerical effects: the spatial
resolution, and consequently $k_{\textrm{max}}$, is smaller and fewer
particles are employed, so that the noise level at small scales is
higher and this slightly affects also the slope.

While the extent of the power-law range at large scales is about a
full decade for all the simulations, the one at sub-ion scales is
usually smaller, being still between half a decade and a decade in
most cases. In this respect, it's important to stress that, although
Fig.~\ref{fig:spectra_singlebeta} provides an insight on the whole
range of $\beta$, including the extreme regimes, only the central
panel is truly representative of most of the simulations in terms of
the extent of power-law ranges.  The other two panels allow
appreciating how the method works in the worst cases, i.e., when the
sub-ion range is reduced due to the shift of the break towards smaller
scales (for low betas) or to the lower resolution (for large betas).
Although the extent of the fit intervals at sub-ion scales is not as
large as a full decade, we can still identify a power-law behavior
rather than an exponential cutoff, which would be typical of resistive
effects. The local spectral index $\alpha_2$ is observed to be
reasonably constant, with only very small variations, in the whole fit
interval in all the panels of Fig.~\ref{fig:spectra_singlebeta},
especially in the middle one.  The same result would not hold in the
case of an exponential cut-off, since $\alpha_2$ would clearly
decrease before starting growing again at small scales due to
numerical noise.
   
Alternatively, we also determined the break position by performing the global
fits over the two ranges of wavevectrors selected with the
method explained above, plotting the straight lines which correspond to
the best fit (blue and red dashed lines in Fig.~\ref{fig:spectra_singlebeta}, 
respectively) and determining the break as the intercept between the two.
We can say a posteriori that the difference between the determination
of the break by the two different methods is almost negligible for all the
simulations performed.

\begin{figure}
\centering
\includegraphics[width=0.47\textwidth]{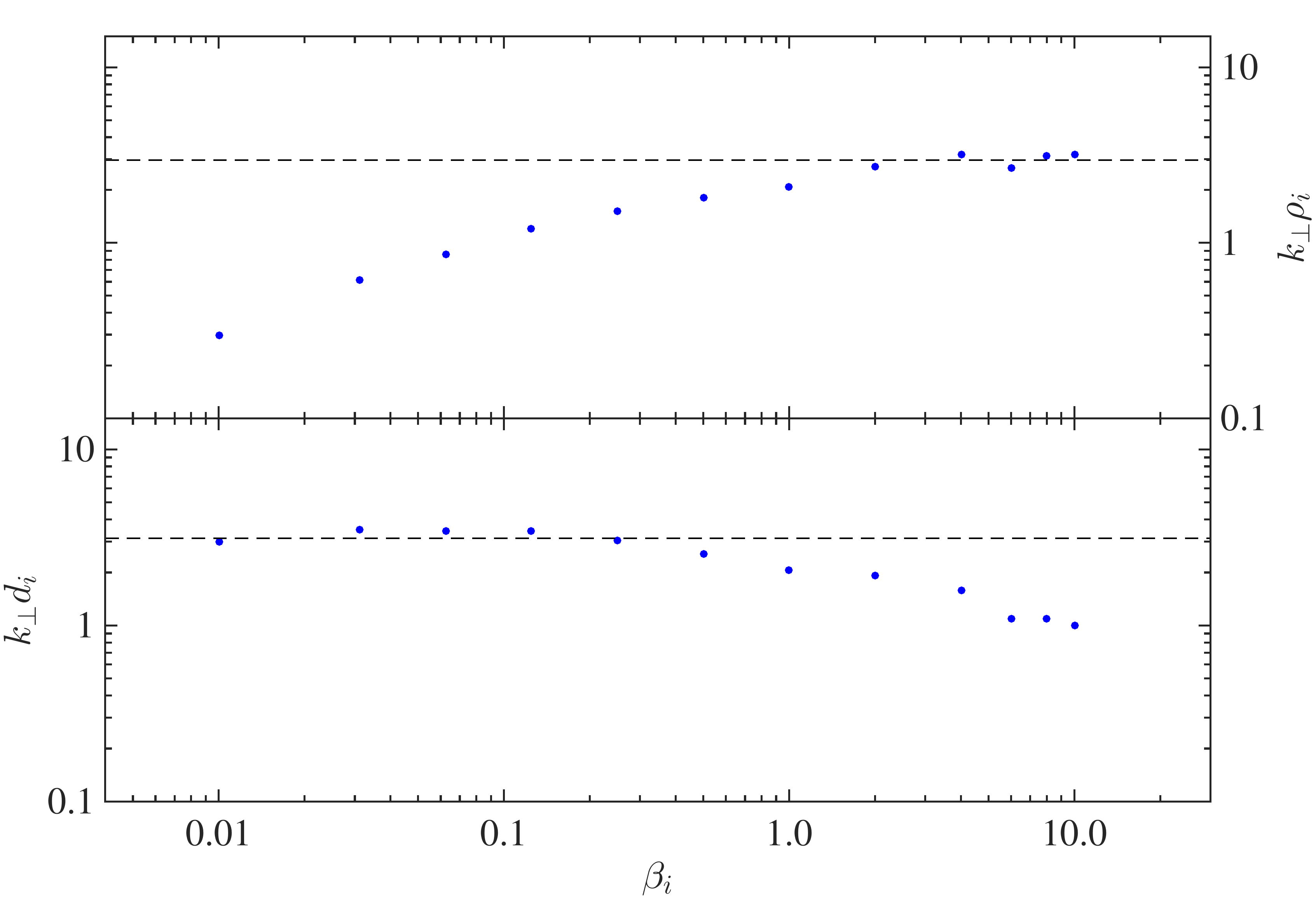}\\ 
\includegraphics[width=0.485\textwidth,trim={1.8cm 0 0 1cm},clip=true]{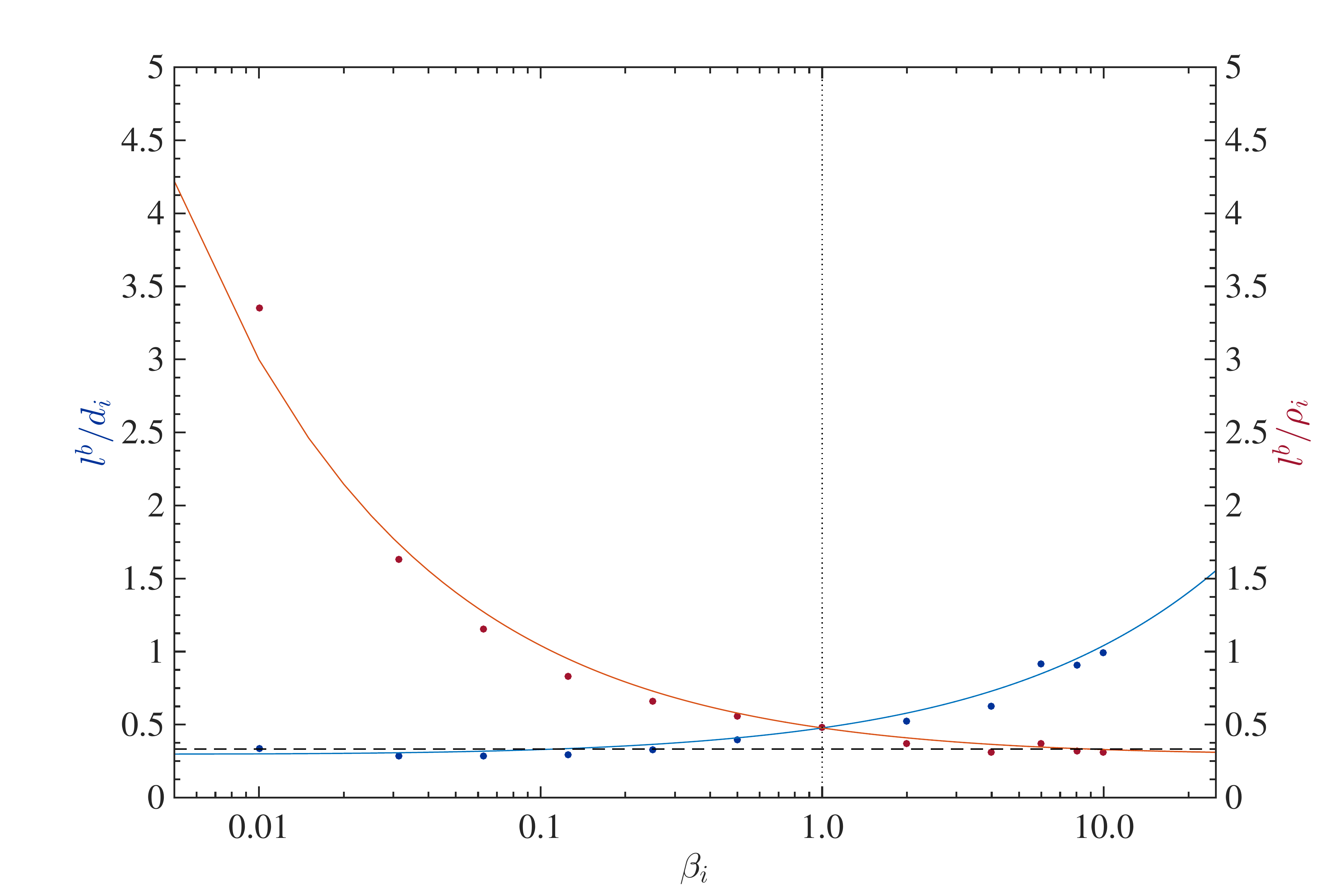}
\caption{Top panel: blue points denote the wavevector $k_\perp^b$
  associated with the spectral break in the magnetic fluctuations,
  normalized to $\rho_i$ (top half) and to $d_i$ (bottom half), as a
  function of the plasma $\beta$ for all the simulations performed.
  Dashed lines show the asymptotic values $k_\perp \rho_i \sim 3$ (top
  half) and $k_\perp d_i \sim 3$ (bottom half).  Bottom panel: blue
  and red points denote the length scale $l^{\textrm{b}}$ of the
  break, normalized to $d_i$ and $\rho_i$, respectively, as a function
  of the plasma $\beta$. A blue and an orange curves represent the
  empirical relation $l^{\textrm{b}} = (d_i + \rho_i - \sqrt{d_i
    \rho_i}/2)/3$, computed in terms of $d_i$ and $\rho_i$,
  respectively.}
\label{fig:break_vs_scales}
\end{figure}

In the top panel of Fig.~\ref{fig:break_vs_scales}, we report the
computed break scale in terms of $k_\perp \rho_i = 1$ (top panel) and
$k_\perp d_i = 1$ (middle panel), as a function of $\beta$, for all
the simulations performed.  For $\beta \gg1 $ the points seem to
settle towards an asymptotic value which is fixed in terms of
$\rho_i$.  By fitting with a straight line, we get $k_\perp \rho_i
\sim 3$.  On the contrary, when the plasma beta decreases to values
$\beta\ll1$ the points seem to approach a constant value in terms of
$d_i$. By fitting with a straight line, we get an asymptotic value
$k_\perp d_i \sim 3$.  Since the ion inertial length and the ion
gyroradius are related by $\rho_i = d_i \sqrt{\beta_\perp}$, we find
that $d_i \gg \rho_i$ for $\beta \ll 1$ and $\rho_i \gg d_i$ for
$\beta \gg 1$. Therefore, the break is found to be related to the
largest of the two scales in both these separated ranges of values of
the plasma beta.  Differently, the spectral break does not show any
clear correlation with one of the two scales when $\beta \sim 1$,
meaning that it is likely related to a combination of $d_i$ and
$\rho_i$ when they are comparable.

In the bottom panel of Fig.~\ref{fig:break_vs_scales}, we report the 
length scales associated to the break versus the plasma beta for
all the simulations performed, rescaled by $d_i$ and by $\rho_i$ 
(red and blue points, respectively).  We have looked for a relation
$l^{\textrm{b}} = l(\beta)$ that could properly mimic the behavior of
the spectral break over the whole range of values of $\beta$ that we
have investigated, i.e., being dimensionally correct, approaching the
two asymptotic values for $\beta\ll1$ and $\beta
\gg 1$, respectively, and passing through $d_i/2 \equiv \rho_i/2$ for
$\beta = 1$.  The relation
\begin{equation}
  l^{\textrm{b}} = \frac{1}{3} \left( d_i + \rho_i - \frac{ \sqrt{d_i
      \, \rho_i} }{2} \right) = \frac{d_i}{3} \left( 1 +
  \beta_\perp^{1/2} - \frac{ \beta_\perp^{1/4} }{2} \right)
\label{eq:breakscale}
\end{equation}
meets all the requirements and seems to represent quite a good
approximation. In the same figure, we plot this analytical expression
for $l^{\textrm{b}}/d_i$ and $l^{\textrm{b}}/\rho_i$ versus the plasma beta 
(blue and orange curves, respectively), while the black dashed line
represents just a reference corresponding to the
two asymptotic values $l^{\textrm{b}}/d_i=l^{\textrm{b}}/\rho_i \sim 1/3$.

\section{Discussion and conclusions}
\label{sec:conclusions}

We investigated the spectral properties of plasma turbulence around
ion scales, by performing 2D high-resolution hybrid particle-in-cell
simulations with different values of the plasma beta from $0.01$ to
$10$.

\begin{figure}
\centering
\includegraphics[width=0.48\textwidth]{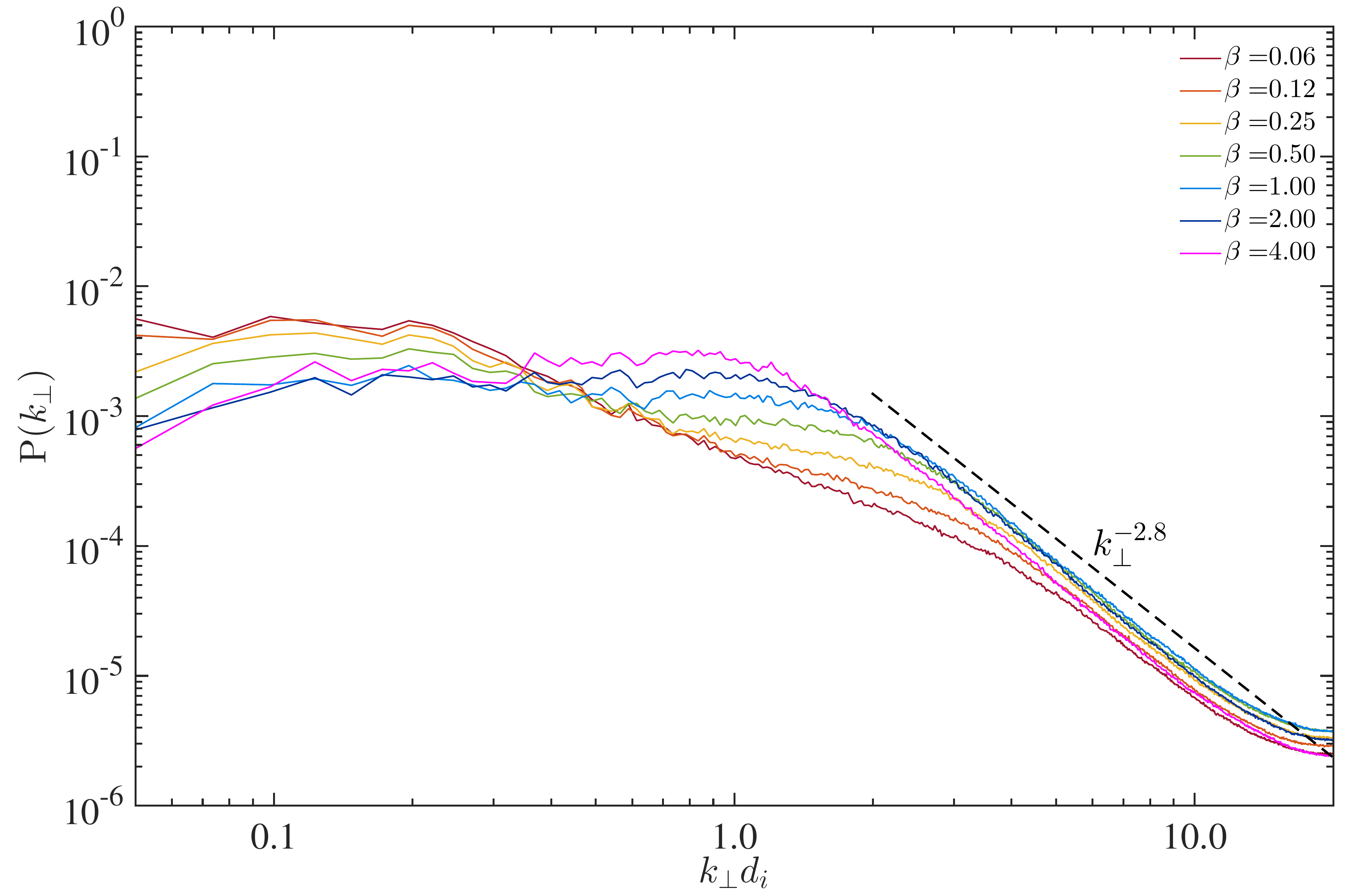}\\ 
\includegraphics[width=0.48\textwidth]{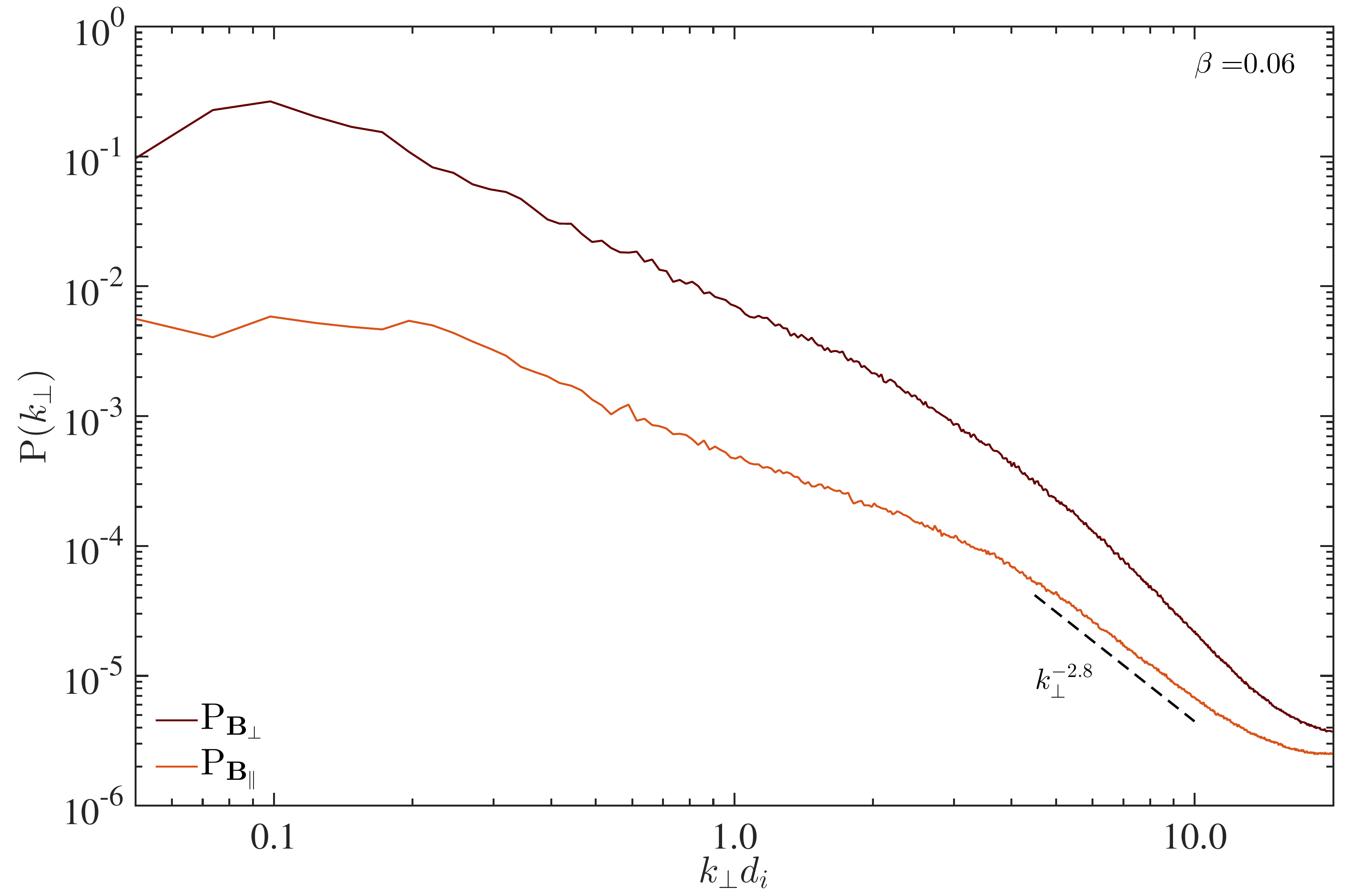}\\ 
\includegraphics[width=0.48\textwidth]{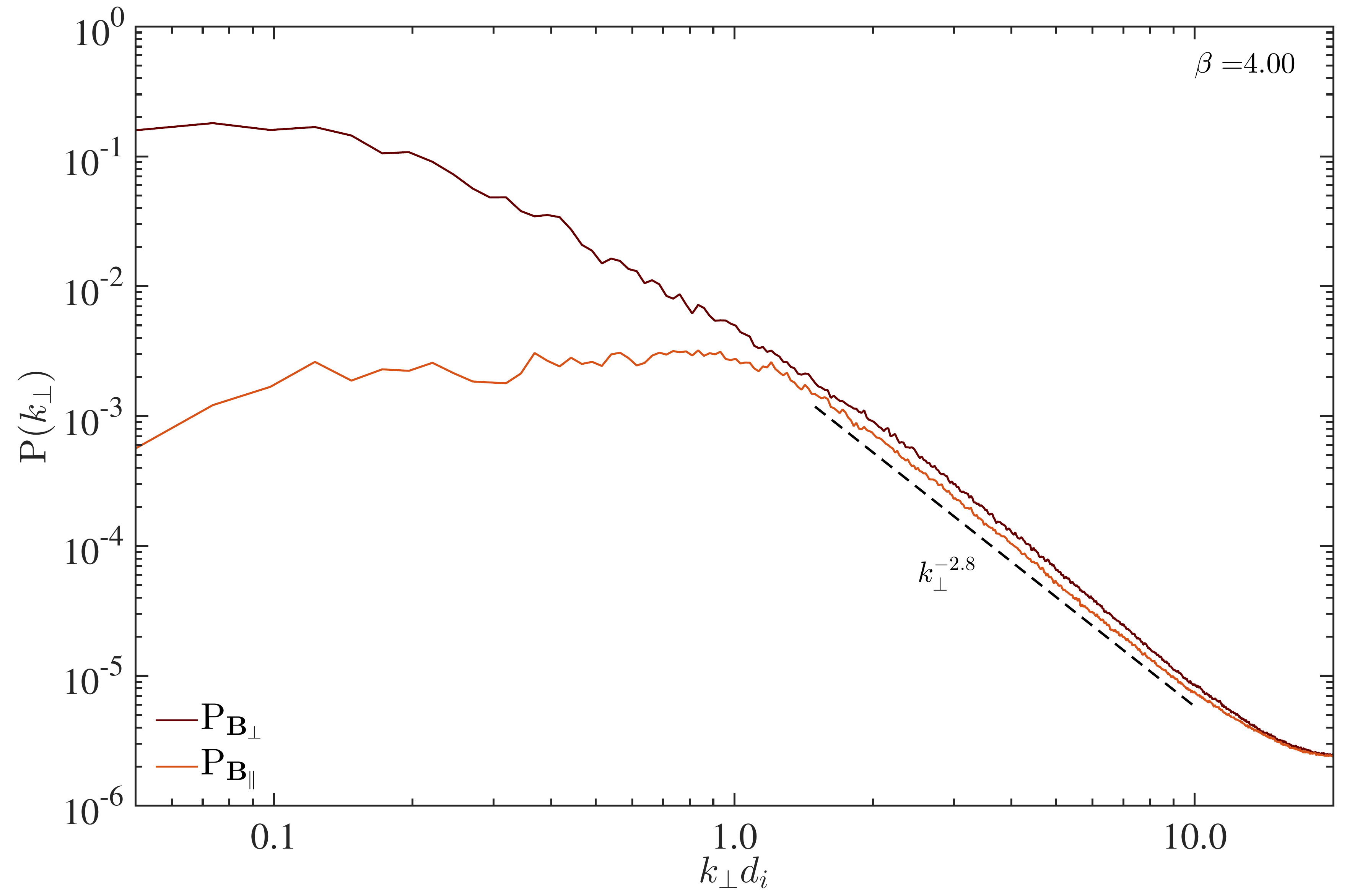}
\caption{Top panel: power spectra of the parallel magnetic
  fluctuations for different values of $\beta$ (for the sake of
  clarity, only the simulations with the same initial setup, i.e., the
  same level of initial fluctuations and the same spatial resolution,
  are shown here).  Middle panel: comparison between the power spectra
  of the perpendicular and the parallel magnetic fluctuations (dark
  red and orange, respectively) for a low-beta case.  Bottom panel:
  the same as in the middle panel, but for a high-beta case. In all
  panels, a power law with spectral index of -2.8 is reported as a
  reference for the scaling at sub-ion scales.}
\label{fig:spectra_bparallel}
\end{figure}

The total magnetic energy spectra exhibit a power-law behavior at
kinetic scales with a slope varying with the plasma $\beta$. A
relatively hard spectrum, with a spectral index of about $-3.6$ for
$\beta \ll 1$, becomes less and less steep as $\beta$ increases,
reaching a value around $-2.9$ when $\beta$ is of order of unity or
higher (the further steepening in the power law observed for $\beta>4$
is likely a numerical artefact due to the lower resolution and the
higher noise level of those simulations).  A similar, quite large
variability of the spectral index of the magnetic field spectrum at
sub-ion scales is also found in solar wind observations, typically
between $-2$ and $-4$ \citep[e.g.,][]{Leamon_al_1998, Smith_al_2006b,
  Alexandrova_al_2008b, Bruno_al_2014}. Such spread is mainly observed
in the kinetic region close to the break, i.e., in a small range of
sub-ion frequencies limited to $f < 10 \; \textrm{Hz}$. Some of this
large variability could be related to the presence of ion instabilities 
or other effects \citep{Hellinger_al_2015, Lion_al_2016}.
However, when the instrumental accuracy allows to further extend the measurement
towards the electron scales, a convergence around $\sim -2.8$ is
rather found \citep[e.g.,][]{Alexandrova_al_2013, Bruno_al_2014}, with
a smaller variability between $-2.5$ and $-3.1$
\citep{Sahraoui_al_2013}.  Indeed, in all our simulations, a more
universal power-law (e.g., independent from $\beta$) is observed for
the parallel magnetic spectrum in the kinetic range, with a spectral
index of $-2.8$. This is clearly shown in the top panel of
Fig.~\ref{fig:spectra_bparallel}, where the power spectra of the
parallel magnetic fluctations are reported for different values of
$\beta$ between $1/16$ and $4$. Such power-law scaling is consistent
with our previous simulations \citep{Franci_al_2015b} and with
observations \citep{Alexandrova_al_2009}.  We speculate that this
different behaviour of the total and parallel magnetic spectra
reflects the different dependence of the compressibility on the plasma
$\beta$ in the inertial and kinetic range: the strong magnetic
compressibility typically observed in the kinetic range
\citep{Alexandrova_al_2008b, Salem_al_2012, Kiyani_al_2013} is reached
in a different way from the inertial range according to its level of
compressibility, i.e., the plasma $\beta$. The middle and bottom panel
of Fig.~\ref{fig:spectra_bparallel} show that, at small scales, the
perpendicular magnetic fluctuations tend to reach asymptotically the
same level as their parallel counterparts. This results in the steeper
power spectrum of the perpendicular (and hence, the total) magnetic
field for low $\beta$, since such coupling is expected to be reached
at scales smaller than the resolved ones. For high $\beta$, the level
of parallel fluctuations is higher in the inertial range, so that the
coupling already occurs at ion scales and the same scaling for the
parallel and perpendicular power spectra is observed.

The shape of the ion-scale transition also depends on $\beta$: it is
quite sharp for high values and smoother for low ones.  The reason of
this different behavior is not clear yet, although it could be related
to the possible different nature of the processes determining the
break in different regimes of $\beta$.

The associated scale-length to
this break is found to be proportional to $d_i$ for $\beta \ll 1$ and
to $\rho_i$ for $\beta \gg 1$, i.e., to the largest of the two in both
limits in good agreement with solar wind turbulence at high and low
beta \citep{Chen_al_2014}.  For intermediate cases, i.e., when $d_i
\sim \rho_i$, a combination of the two better reproduces the scaling
with $\beta$ observed in our simulations.
Different processes can be invoked in order to explain the position of
the inertial-kinetic transition and the shape of the magnetic power
spectrum at sub-ion scales. Landau damping has been considered
relevant for the steepening and in introducing a non-universal power
law in the magnetic spectrum \citep[e.g.,][]{Howes_al_2011,
  Passot_Sulem_2015,Sulem_al_2016}. However, in our study, the main
drivers of the Landau damping (i.e., the electrons) are not treated
kinetically. Alfv\'en waves resonances can determine the scale where
the magnetic power spectrum steepens 
\citep[e.g.,][]{Gary_Borovsky_2004,Bruno_Trenchi_2014,Bruno_al_2014}. However,
cyclotron damping requires a significant contribution of
$k_\parallel$, which is strongly inhibited in our simulations by the
2D geometry, although a local propagation of modes with $k_\parallel
\neq 0$ can occur through the local bending of the magnetic field
lines \citep[e.g.,][]{Hellinger_al_2015}.

The transition from shear Alfv\'en waves to kinetic Alfv\'en waves
(KAW) represents a possible explanation for the ion-scale break, at
least when $\beta \gg 1$. This fact is corroborated by the
polarizations of the fluctuations at small scales (not shown here, but
the particular case with $\beta = 0.5$ was already presented in
\citet{Franci_al_2015a}), which show a good agreement with the
prediction of the KAW linear theory for $\beta \gtrsim 1$ and are
consistent with the fact that $\rho_i$ is the expected scale for such
transition in this regime \citep[e.g.,][]{Chen_al_2014}. This would
be consistent with the results by \citep{Cerri_al_2016}, which observe
a dominance of KAW for $\beta \gtrsim 1$ but not for lower betas.

The dispersive nature of KAW in regulating the break in the magnetic
field spectrum is much more problematic when $\beta\ll 1$ since $d_i$,
the scale we observe in this limit, seems to be relevant for KAW only
under special circumstances ($T_i \ll T_e$ and for $\beta_e \gg 1$) or
in presence of a large component of turbulence in $k_\parallel$
\citep{Chen_al_2014}. These conditions are not fulfilled in our simulations.

It has been suggested that the ion-scale transition could be mainly
due to the dissipation occurring in reconnecting current
sheets. Indeed, the scale at which such transition occurs corresponds
to the maximum in the current density spectrum, suggesting that most
of current structures develops at that scale.  As already shown in
Fig.~2 of \citep{Franci_al_2015b} and therein discussed, while
turbulence develops many current sheets are generated around and
between coherent structures. Once formed, these are observed to
quickly disrupt due to the onset of fast reconnection.  A look at the
out-of-plane current density in our simulations seems to qualitatively
support this interpretation: the current sheets form and shrink, and
their width when reconnection occurs seems to be of the order of $d_i$
in all simulations with $\beta < 1$ and larger when $\beta > 1$.  If
this process was the main responsible for the break, we would expect
the associated length scale to be related to the current sheet
width. Solar wind observations \citep{Leamon_al_2000, Vasquez_al_2007,
  Borovsky_Podesta_2015} indicate that such width, although variable,
scales better with $d_i$ for $\beta < 0.1$ and with $\rho_i$ for
$\beta > 4$ \citep{Vasquez_al_2007}.  Actually, the agreement between
our numerical results and \citep{Vasquez_al_2007} observations also
extends to large betas. This could be a hint that the break might be
related to reconnection for all betas. Current sheets and
reconnection likely play an important role in plasma turbulence 
\cite[cf.,][and references therein]{Servidio_al_2015}.

\citet{Chen_al_2014} pointed out that this scaling is in
contradiction with results from previous simulations
\citep{Cassak_al_2007} and laboratory measurements of reconnection
with a large guide field \citep{Egedal_al_2007}, where the current
sheets thickness in the $\beta \ll 1$ condition is found to be the
sound gyroradius, $\rho_s=\sqrt{T_{\rm e}/T_{\rm i}}\rho_{\rm i}$
($\equiv \rho_i$ when $T_e = T_i$). However, the definitions of
$\rho_s$ used in the above papers can not be easy exploited in solar
wind observations, since they take into account only the reconnecting 
(i.e., in-plane) magnetic field.

In our simulations, the scale at which the magnetic field spectrum
breaks is found to be quite well approximated by a single relation,
$l^\textrm{b} = l(\beta_\perp)$, Eq.~(\ref{eq:breakscale}), being able
to recover both the asymptotic behavior in the limits of low
($l^\textrm{b} \propto d_i$) and high beta ($l^\textrm{b} \propto
\rho_i$) and the intermediate-regime scaling (a combination of the
two). This relationship is qualitatively similar to that proposed by
\citet{Bruno_Trenchi_2014} (basically, a mean of the $d_i$ and
$\rho_i$) for values $\beta\sim 1$, although here it can not be easily
interpreted in terms of a resonant condition.  At this level, it
should be regarded as an empirical relation which can mask either a
single process dominating for all betas (e.g., the current sheet
width) or different processes, each one dominating at one
characteristic ion scale when they are well separated and, instead,
mixing in the intermediate regime \citep{Markovskii_al_2008}, for
example kinetic Alfv\'en waves for high $\beta$ and magnetosonic-like
for low $\beta$ \citep[e.g.,][]{Cerri_al_2016}.

The simulation method used in this work has a couple of limitations,
i.e., the lack of electron kinetic processes and the reduced
dimensionality.  In the hybrid approach electrons are treated as a
fluid, thus not capturing processes such as the electron Landau
damping and electron kinetic instabilities. Although these processes
may affect the turbulent dynamics at very small scales, possibly
modifying the spectral properties in the sub-ion range, they are not
expected to change the transition from large- to small-scale
turbulence at ion-scales. The dissipation at the electron scales is to
some extent replaced by using a finite resistivity, $\eta$.  In
\citet{Franci_al_2015b}, we qualitatively checked if and how the slope
in the sub-ion range is affected by $\eta$.  We showed that by
fine-tuning its value and controlling the scale associated to it, one
can be able to separate regimes where the sub-ion spectral behavior
reflects a physical cascade from cases where the change in the slope
can be ascribed to purely resistive effects. Therefore, in the present
work we can be reasonably confident that the sub-ion spectral slopes
(where shown and discussed) are physically meaningful and indicative
of a cascade process. Consistently with this, in the series of runs
3--9 we only vary $\beta$ by keeping the same level of fluctuations,
resolution, and resistivity, so we can exclude that steeper slopes for
lower betas originate from more and more effective resistive term and
claim that the systematic change in $\alpha_2$ with $\beta$ has a
physical motivation.

The used 2D geometry allows the very high resolution and the large
simulation box size needed to accurately determine the position of the
ion-scale break, but the reduced dimensionality might affect the
development of the turbulent cascade and of kinetic
instabilities. However, preliminary 3D runs in a similar settings
confirm that the perpendicular cascade is not strongly modified (see
also \citet{Servidio_al_2015}), thus suggesting that 2D simulations
represent an adequate tool to investigate the spectral break.

In conclusion, our main findings about the effects of $\beta$ on the
magnetic field spectrum are that: i) the slope of the power law in the
sub-ion range depends on $\beta$, ii) the shape and position of the
ion-scale transition also depend on $\beta$, and iii) we inferred
an empirical relation for the length corresponding to the ion spectral
break, $l^{\textrm{b}} = [ d_i + \rho_i - (d_i \rho_i)^{1/2}/2] /3$,
that well describes the simulation results for all values of $\beta$.

Further investigation is needed to better clarify the nature of the
ion-scale spectral break. An accurate statistical study about the
current sheets thickness \citep[e.g.,][]{Servidio_al_2009} would allow
to quantitatively investigate its scaling with the plasma beta.  An
analysis about the effects of $\beta$ on the ion heating and
temperature anisotropy and their correlation with the current density
and vorticity \citep{Servidio_al_2015, Franci_al_2016} will be also
the subject of future work.  High-resolution 3D simulations are
necessary in order to extend and validate the present results.

\acknowledgments

LF is funded by the Ente Cassa di Risparmio di Firenze. PH
acknowledges GACR grant 15-10057S. We acknowledge PRACE for awarding
us access to resource Cartesius based in the Netherlands at SURFsara
through the DECI-13 (Distributed European Computing Initiative) call
(project HybTurb3D), and the CINECA award under the ISCRA initiative,
for the availability of high performance computing resources and
support (grants HP10C877C4 and HP10BUUOJM). Data deposit and
preservation through a persistent identifier (pid) were provided by
the EUDAT infrastructure (\url{https://www.eudat.eu/}).

\bibliographystyle{apj-eid}

\end{document}